\documentclass[aps,twocolumn,showpacs,amsmath,amssymb]{revtex4}
\usepackage{graphicx}
\usepackage{dcolumn}
\usepackage{bm}

\newcommand{\br}{{\bf r}}
\newcommand{\bp}{{\bf p}}
\newcommand{\bK}{{\bf K}}
\newcommand{\bA}{{\bf A}}
\newcommand{\Area}{{\cal  A}}

\newcommand{\fr}{\frac}

\newcommand{\Identity}{\mbox{.\bfseries \large 1}}
\newcommand{\Identityy}{\mbox{\bfseries \large 1}}

\newcommand{\smpl}{\! + \!}
\newcommand{\smmi}{\! - \!}

\newcommand{\vf}{v_{\scriptscriptstyle F}}

\begin{document}

\title{Berry phase in graphene: a semiclassical  perspective} 
\author{Pierre Carmier and Denis Ullmo}
\affiliation{Univ.\ Paris-Sud, LPTMS UMR 8626, 91405
  Orsay Cedex, France}
\affiliation{CNRS,  LPTMS UMR 8626, 91405   Orsay Cedex, France}    
\date{\today}

\begin{abstract}

  We derive a semiclassical expression for the Green's function in
  graphene, in which the presence of a semiclassical phase is made
  apparent.  The relationship between this semiclassical phase and the
  adiabatic Berry phase, usually referred to in this context, is
  discussed. These phases coincide for the perfectly linear Dirac
  dispersion relation.  They differ however when a gap is opened at
  the Dirac point. We furthermore present several applications of our
  semiclassical formalism.  In particular we provide, for various
  configurations, a semiclassical derivation of the electron's Landau
  levels, illustrating the role of the semiclassical ``Berry-like'' phase.

\end{abstract}

\maketitle

\section{Introduction}

Graphene \cite{Novoselov05,CastroNeto07}, a two-dimensional carbon
based material forming a honeycomb lattice, has attracted a lot of
attention since its experimental isolation has been proved possible
\cite{Novoselov04,Berger04}. It is a gapless
semiconductor in which, near half filling, electrons behave like
massless Dirac particles, obeying a linear dispersion relation.  Among
the unusual properties of this two-dimensional carbon material stand
out very distinctive quantum Hall properties, and in  particular the
$\sqrt{n}$ dependence  of the energy in terms of the Landau level number $n$,
and the existence of a Landau level with zero energy, which is
associated with the presence of a Berry phase \cite{Novoselov05,Zhang05}.

The existence of this Berry phase and its implications for the Landau
levels have been discussed in many places in different contexts (see
e.g.\ \cite{Ando98,Mikitik98,Zhang05}).  The direct connection between
the Berry phase and the observable quantities under discussion is
however not always as transparent as one may wish, and situations
where, either because of disorder, or because one would like to
confine the electrons into a finite region of space, a position
dependent electrostatic potential or mass term is introduced, are
usually not addressed.

The aim in this paper is to revisit this question of Berry phase in
graphene within a semiclassical, and more specifically semiclassical
Green's function, perspective.  For sake of clarity, our emphasis in
this present work will be more in providing this new point of view,
and we shall therefore mainly illustrate it with the discussion of the
standard problem of the Landau levels of electrons in a perpendicular
and uniform magnetic field.  Even in this familiar framework, we shall
see however that our semiclassical approach makes it possible to
address some non-trivial questions, such as the role of the Berry
phase in situations for which a small mass term has to be included,
opening in this way a gap at the Dirac point.

This article is therefore organized as follows.  In
section~\ref{sec:deriv}, we derive, following closely the formalism of
Bolte and Keppeler \cite{Bolte99}, the expression for the
semiclassical Green's function in graphene.  In particular we discuss
in details the origin of the term corresponding to the Berry phase.
These results are extended in section~\ref{sec:bilay&GTF} to a
bilayer of graphene.  We furthermore provide both for the monolayer
and the bilayer cases the expression of the Gutzwiller trace formula
for the semiclassical density of states, valid when classical periodic
orbits are isolated.  As an illustration of the Green's function
formalism, we then apply it in section~\ref{sec:landau} to the
computation of Landau levels for a graphene sheet in constant magnetic
field. We will see in particular that the modifications brought in by,
for instance, trigonal warping, are easily included within our
semiclassical formalism.  We then come back in
section~\ref{sec:1/2classvsquant} to the discussion of the
relationship between the semiclassical ``Berry-like'' phase obtained
in our approach and the adiabatic Berry phase \cite{Berry84} usually
discussed in this context.

\section{Semiclassical Green's function for graphene}
\label{sec:deriv}

Starting from a tight-binding nearest neighbor model, the graphene
Hamiltonian at low energies can be obtained by expanding the momentum
near the Dirac points $\bK$ and $\bK'$ of the Brillouin zone. For pure
graphene, one obtains in this way in momentum representation
\cite{Wallace47,Slonczewski58,McClure56,Goerbig06}
\begin{equation} \label{eq:Hog}
{\cal H}^0_g = \vf(\alpha \sigma_x p_x + \sigma_y p_y) =
\vf\begin{pmatrix} 0 & {\alpha} p_x - i p_y \\  {\alpha} p_x + i p_y & 0
  \\ \end{pmatrix} \; ,
\end{equation}
where the matrix structure originates from the existence of two
sub-lattices (denoted $A$ and $B$ below) in the graphene honeycomb
structure.  In this equation, $\vf = {3ta}/({2\hbar})$ is the Fermi
velocity, with $t$ the hopping parameter and $a$ the lattice constant,
$\alpha$ is the valley index ($\alpha = \pm 1$) labelling the two
inequivalent points $\bK$ and $\bK'$ in the Brillouin zone (not to be
confused with the sub-lattice index), $\bp$ is the momentum measured
from these points, and $\sigma_{x,y}$ are Pauli matrices.  This linear
approximation to the graphene Hamiltonian will be valid as long as the
condition $|\bp| \ll {\hbar}/{a}$ is fulfilled.

We are interested here in a more general situation than the one of
pure graphene, and would like to consider the case where, because of
either disorder or the need to confine the electrons in some part of
the graphene sheet, an electrostatic potential $U(\br)$ and/or a
[possibly position dependent] mass $m(\br)$ have to be taken into
account. We will not consider however tunneling contributions related
to the Klein paradox, or boundary effects that may occur at the
(zigzag, armchair, or generic) edges of the graphene sample. The
graphene Hamiltonian then takes the more general form
\begin{equation}
\label{eq:hamgen}
{\cal H}_g  = \vf(\alpha \sigma_x \hat \Pi_x + \sigma_y \hat \Pi_y) +
U(\br)
\Identity_2 + m(\br)\vf^2\sigma_z  
\; ,
\end{equation}
in which the magnetic field ${\bf B}(\br) = {\bf \nabla} \times
\bA(\br)$ (if any) is taken into account by the Peierls substitution
\begin{equation}
\hat \bp \to \hat {\bf \Pi} = \hat \bp + e \bA (\br) \; , 
\end{equation}
with $\bA(\br)$ the vector potential and $\hat \bp \equiv
-i\hbar\fr{\partial}{\partial \br}$.

For this problem, the Green's function $G(\br'',\br')$ is actually
a $2 \times 2$ matrix defined by the differential equation 
\begin{equation}
\label{eq:diff}
(E\Identity_2 - {\cal H}_g)G(\br'',\br';E) = \delta(\br'' - \br')\Identity_2
\end{equation}
(where ${\cal H}_g$ is applied to the variable $\br''$).  To obtain a
semiclassical solution of this equation, we shall proceed in two steps.
 First, assuming $\br''$ is far from the  source location $\br'$, we solve
semiclassically (i.e. in the WKB approximation) the 
Schr\"odinger equation 
\begin{equation} \label{eq:schroe}
(E\Identity_2 -{\cal H}_g)G=0 \; .
\end{equation}
In a second stage we match this general solution to the exact Green's
function of the ``free'' (i.e.\ with constant potential and mass)
problem, valid near the singularity $\br'$.  We proceed now with this
derivation.

\subsection{Far from the singularity: the WKB approximation}

Following \cite{Bolte99}, we seek a semiclassical solution of
eq.~(\ref{eq:diff}) with $G$ of the form 
\begin{equation}
\label{eq:semiGr}
G(\br'',\br';E) = \Gamma(\br'',\br')
\exp\left[ {\fr{i}{\hbar}S(\br'',\br')} \right], 
\end{equation}
where $\Gamma$ is a 2x2 matrix.  To lighten the notation, we drop for
now the explicit dependence in the source position $\br'$.  Inserting
(\ref{eq:semiGr}) into (\ref{eq:schroe}) and expanding in $\hbar$ the
resulting expression, we obtain at order $O(\hbar^0)$
\begin{equation}
\left( E\Identity_2 - H(\fr{\partial S}{\partial \br''},\br'') \right)
\Gamma(\br'')  =   0 \; , \label{eq:order0}
\end{equation}
and at order
$O(\hbar^1)$ 
\begin{equation}
  \frac{\partial H}{\partial \bp} \cdot \frac{\partial }{\partial
    \br''}\Gamma(\br'') 
  = 
  \vf  (\alpha\sigma_x\fr{\partial}{\partial x''} +
  \sigma_y\fr{\partial}{\partial y''})\Gamma(\br'') 
  =   0 \; ,
 \label{eq:order1} 
\end{equation}
where $H(\bp,\br)$ is the classical symbol associated with the quantum
Hamiltonian ${\cal H}_g$. 

This classical Hamiltonian can be diagonalized, with the eigenvalues 
\begin{equation} \label{eq:Hclass}
H^{\pm}(\bp,\br)=U(\br) \pm\sqrt{m^2(\br)\vf^4+\vf^2{\bf \Pi}^2}
\end{equation}
and the corresponding normalized eigenvectors $V^{\pm}(\bp,\br)$ 
(whose explicit expressions are given in
appendix~\ref{sec:appA}).  Writing the matrix $\Gamma(\br'')$ as
$[V^{\pm}(\frac{\partial S}{\partial \br''},\br'') \cdot{\tilde
  \Gamma}^{\pm}(\br'') ]$, with ${\tilde \Gamma}^{\pm}$ a $1 \times 2$
matrix, the order $\hbar^0$ equation becomes
\begin{equation}
\label{eq:HJchar}
E-H^{\pm}(\fr{\partial S}{\partial \br''},\br'')=0 \; ,
\end{equation}
where the $\pm$ sign must be taken according to the sign of
$E-U(\br'')$.

Eq.~(\ref{eq:HJchar}) is the usual scalar Hamilton-Jacobi equation,
which can be solved by the method of characteristics \cite{Maslov81}.
This amounts to constructing a 2-dimensional Lagrangian manifold
${\cal L}$ (in the 3-dimensional energy surface in phase space)
built as a 1-parameter family of trajectories following the classical
equations of motion
\begin{eqnarray*}
\dot { \br} & = & \fr{\partial H^{\pm}}{\partial \bp}(\bp,\br)
\; , \\
\dot { \bp} & = &  -\fr{\partial   H^{\pm}}{\partial
  \br}(\bp,\br) \; .
\end{eqnarray*}  
Given any such manifold, the action $S(\br'') = \int^{\br''} \bp d \br$,
where the integral is taken on an arbitrary path on ${\cal L}$, is a
solution of (\ref{eq:HJchar}).

The specific Lagrangian manifold that will correspond to the proper boundary
conditions for $G(\br'',\br')$ near the source $\br'$ is the one
  obtained from the 
trajectories leaving $\br'$ with an arbitrary initial momentum $\bp'$
at energy $E$: 
\begin{equation} \label{eq:manif}
\begin{split}
{\cal L}^{\pm} = & \{ ( \bp(t),  \br(t)), t \in [0,\infty), \\
& \mbox{such that }  \br(0) = \br', \,  H^{\pm}(\bp(0), \br(0))=E \} 
\end{split}
\end{equation}
(each point on the manifold is therefore parameterized by the time  $t$
and the initial momentum $ \bp(0)$). The corresponding action
can then be expressed as 
\begin{equation} \label{eq:Sgreen}
S^{\pm}(\br'',\br') = \int^{\br''}_{\br'}   \bp  \cdot \dot { \br} \, dt 
\end{equation}
along a trajectory $(\bp(t),\br(t))$ joining $\br'$ to
$\br''$ at energy $E$. 

Having obtained a solution of the $O(\hbar^0)$ equation, the prefactor
$\tilde \Gamma$ is then determined by the $O(\hbar^1)$
equation~(\ref{eq:order1}), which, after multiplication on the left by
$V^{\pm\dagger}(\fr{\partial S^{\pm}}{\partial \br''},\br'')$, can be
expressed as $\Box {\tilde \Gamma}^{\pm} = 0$, where
\begin{equation*}
\Box \equiv \left( V^{\pm\dagger}(\fr{\partial S^{\pm}}{\partial
    \br''},\br'') \fr{\partial H}{\partial \bp}.\fr{\partial
  }{\partial \br''} \right) V^{\pm}(\fr{\partial S^{\pm}}{\partial
  \br''},\br'') \; .
\end{equation*}
The operator $\Box$ can be decomposed as $\Box = \Box_{(1)} +
\Box_{(2)}$ with
\begin{equation} \label{eq:Box1}
\Box_{(1)} = \left( V^{\pm\dagger}\fr{\partial H}{\partial
    \bp}V^{\pm} \right) .\fr{\partial }{\partial \br''} 
\end{equation}
and 
\begin{equation} \label{eq:Box2}
\Box_{(2)} = V^{\pm\dagger}\fr{\partial H}{\partial
  \bp}. \left( \fr{\partial V^{\pm}}{\partial \br''} \right) \; .
\end{equation}

Noting that first order perturbation theory implies $V^{\pm\dagger}
({\partial H}/{\partial \bp}) V^{\pm} = ({\partial H^{\pm}}/{\partial
\bp}) $, one has straightforwardly that 
\begin{equation}
\Box_{(1)} = \fr{\partial H^{\pm}}{\partial \bp}.\fr{\partial }{\partial \br''}
\end{equation}
and that 
\begin{equation}
{\rm Re}(\Box_{(2)}) = \fr{1}{2}\fr{\partial }{\partial \br''}. \left(
  V^{\pm\dagger}\fr{\partial H} {\partial \bp}V^{\pm} \right)
= \fr{1}{2}\fr{\partial }{\partial \br''}.\fr{\partial H^{\pm}}{\partial
  \bp} \; .
\end{equation}
(Note here that with respect to spatial derivation, $H^\pm \equiv H^\pm(\br'') =
H^\pm((\partial S^{\pm}/\partial \br'') ,\br'')$).  One recovers in this way, for
the real part of $\Box$, the usual expression valid for a scalar
quantum system \cite{Maslov81}, which is expected since it basically
expresses the conservation of probability.
 
The imaginary part of $(\Box_{(2)})$ however is not constrained by
such a conservation law, as it affects only the phase of $\tilde
\Gamma$, but encodes information about the adiabatic variation of the
eigenvector $V^\pm$ along the followed trajectory.  It needs therefore to be
computed from the explicit expressions of the eigenvector and
eigenvalues of $H(\bp,\br)$.  The details of the algebra are given in
appendix~\ref{sec:appA}. One obtains
\begin{equation}
\label{eq:BK}
\left( \fr{\partial H^{\pm}}{\partial \bp}.\fr{\partial}{\partial \br''} +
  \fr{1}{2}\fr{\partial}{\partial \br''}.\fr{\partial H^{\pm}}{\partial
    \bp} + iM^{\pm} \right)
\tilde{\Gamma}^{\pm} = 0 
\end{equation} 
with
\begin{equation} \label{eq:Mmono}
\begin{split}
M^{\pm}= & \fr{ \alpha \vf^2}{2(E-U(\br''))}  \Big( e{\bf B}+ \\
&  \fr{{\bf
      \Pi}\times \fr{\partial}{\partial
      \br''}(m(\br'')\vf^2-U(\br''))}{m(\br'')\vf^2+E-U(\br'')} \Big) .{\bf
  e}_z
\end{split}
\end{equation}
(${\bf   e}_z$ is the unit vector in the direction perpendicular to
the graphene sheet).

In the absence of the complex term $iM$, the scalar transport equation
$\left( \fr{\partial H^{\pm}}{\partial \bp}.\fr{\partial}{\partial \br''} +
  \fr{1}{2}\fr{\partial}{\partial \br''}.\fr{\partial H^{\pm}}{\partial
    \bp} \right) {\gamma}^{\pm} = 0$ has the usual solution \cite{Maslov81} 
\begin{eqnarray}
\gamma^{\pm} & = & C\fr{\exp({-i\fr{\pi}{2}\mu^{\pm}})}{\sqrt{|{\text
      J}^{\pm}(\br'',\br')|}} \\
J^{\pm}(\br'',\br') & = & -{\dot r''}_{\|} {\dot r'}_{\|} \left(
  \fr{\partial^2 S^{\pm}}{\partial r''_{\bot} \partial
      r'_{\bot}} \right)^{-1}  \nonumber \\
& = & {\dot r''}_{\|} {\dot r'}_{\|} \left(
  \fr{\partial r''_\bot }{\partial p'_{\bot}} \right) \label{eq:J}
\end{eqnarray}
where $r_{\|}$ and $r_{\bot}$ are the coordinates parallel and transverse
to the trajectory (actually Eq.~(\ref{eq:J}) remains valid for
any system of coordinates) and $\mu^{\pm}$ is the Maslov index counting the
(algebraic) number of caustic points.  Writing 
\begin{equation*}
\tilde{\Gamma}^{\pm} = \gamma^{\pm}\Sigma^{\pm}
\end{equation*} 
we obtain that 
\begin{equation*}
(\fr{\partial H^{\pm}}{\partial \bp}.\fr{\partial}{\partial \br''} +
iM^{\pm})\Sigma^{\pm} \equiv (\fr{d}{dt} + iM^{\pm})\Sigma^{\pm} = 0  
\end{equation*}
and therefore $\Sigma^{\pm}(t)  =  \exp\left(i \xi_{\rm sc}
\right) \, \Sigma^{\pm}(t=0)$, with 
\begin{equation} \label{eq:gammasc}
 \xi_{\rm sc}  =  - \int_{0}^{t}M^{\pm}( \bp(t') \; ,
  \br(t'))dt' \; .
\end{equation}
Summing the contributions corresponding to different orbits
$j$ joining $\br'$ to $\br''$ we get
\begin{equation}
\label{eq:Greensemi}
\begin{split}
G(\br'',& \br';E) = \sum_{j:\br' \to \br''} \gamma^{\pm}_j
V^{\pm}_j(\br'') \Sigma_j^{\pm}(t=0) \\
& \exp\left( \fr{i}{\hbar}S^{\pm}_j(\br'',\br') -
    i  \int_{0}^{t_j} M^{\pm}_j( \bp(t') \; , \br(t'))dt' 
\right) \; ,
\end{split}
\end{equation}
where $V^{\pm}_j(\br'') \equiv V^{\pm}( {\partial S^{\pm}_j}/{\partial
  \br''},\br'')$ (and therefore depends not only on $\br''$ but also on
the final momentum $\bp''_j$ of the trajectory $j$).

The semiclassical phase $\xi_{\rm sc}$ Eq.~(\ref{eq:gammasc}) is the
analog, in our context, of a Berry phase \cite{Berry84}.  In the same
way, it has its origin in the adiabatic change of the eigenvectors of
the ``internal degree of freedom'' Hamiltonian $H(\bp(\br),\br)$ along
the classical paths contributing to the semiclassical Green's
function.  Furthermore, in some circumstances, $\xi_{\rm sc}$ {\em
  exactly} corresponds to the genuine Berry phase $\xi_{\rm ad}$
defined for the adiabatic motion along the trajectory.  This will be
the case in particular for ``pure'' (i.e. without mass term) graphene.
In general, however, $\xi_{\rm sc}$ and $\xi_{\rm ad}$ differ
\cite{Littlejohn91prl,Littlejohn91pra}.  We will come back to this
point in section~\ref{sec:1/2classvsquant}, and in particular clarify
the question of which of the two phases is relevant for the Landau
levels.

\subsection{Matching to the exact solution near the source}

Sufficiently close, on the classical scale, to the source $\br'$, we
can neglect the variation of the various potentials and of the mass,
i.e. assume $U(\br)=U_0, m(\br)=m_0$ and $\bA(\br)=0$.  In this  case
we have the expression for the exact retarded Green's function:
\begin{equation*}
G = \begin{pmatrix} G_{AA} & G_{AB} \\ G_{BA} & G_{BB} \end{pmatrix} 
\end{equation*}
with 
\begin{widetext}
\begin{eqnarray}
G_{AA}(\br'',\br',E+i\epsilon) & =  & \left( -i\fr{m_0\vf^2 +  \sqrt{\zeta^2 +
      m_0^2\vf^4}}{4({\hbar}\vf)^2} \right) 
H_0(\fr{\zeta}{{\hbar}\vf}|\br'' - \br'|)  \; ,
\\
G_{AB}(\br'',\br',E+i\epsilon) &= &\left( \alpha\fr{\zeta e^{-i\alpha\phi}}{4({\hbar}\vf)^2}
\right) H_1(\fr{\zeta}{{\hbar}\vf}|\br'' - \br|) 
\end{eqnarray}
\end{widetext}
and $G_{BB} = G_{AA}(m_0 \to -m_0)$, $G_{BA} = G_{AB}(\phi \to -\phi)$.
Here $\zeta=\sqrt{(E+i\epsilon-U_0)^2-m_0^2\vf^4}$, $\phi$ is the phase of 
$p_x + ip_y$ and $H_0$ and $H_1$ are Hankel functions of order 0 and 1. 
Asymptotically, as $|\br'' - \br'| \to +\infty$, 
$G_{AA}$ and $G_{AB}$ take the form
\begin{equation} \label{eq:GaaAss}
G_{AA} \simeq -i\fr{m_0\vf^2 + E -
  U_0}{4({\hbar}\vf)^2}\sqrt{\fr{2}{\pi}}\fr{e^{i(k|\br'' -
    \br'|-\fr{\pi}{4})}}{\sqrt{k|\br'' - \br'|}} 
\end{equation}
\begin{equation}\label{eq:GabAss}
G_{AB} \simeq -i\alpha{e^{-i\alpha\phi}}\fr{\sqrt{(E - U_0)^2 -
    m_0^2\vf^4}}{4({\hbar}\vf)^2}\sqrt{\fr{2}{\pi}}\fr{e^{i(k|\br'' -
    \br'|-\fr{\pi}{4})}}{\sqrt{k|\br'' - \br'|}} 
\end{equation}
with $\hbar k = \fr{1}{\vf}\sqrt{(E-U_0)^2-m_0^2\vf^4} = |\bp|$.

Let us assume $E-U_0 \ge 0$, so that semiclassically we consider the
positive eigenspace $H^{+}$. We note first that, in the free case
considered here, the choice of the
Lagrangian manifold ${\cal L}^+$ given by (\ref{eq:manif}) corresponds
to the action $S^{+}(\br'',\br') = |\bp| \cdot |\br''-\br'|$ and to  
\begin{equation*}
J^{+}(\br'',\br' ) =
\fr{\vf^4}{(E-U(\br'))^2}{|\bp|}.|\br''-\br'| \; ,
\end{equation*} 
so that, as anticipated, the expression (\ref{eq:Greensemi}) matches
the asymptotic expressions (\ref{eq:GaaAss})-(\ref{eq:GabAss}),
provided one chooses $C =\fr{1}{\sqrt{2i\pi\hbar}}
  \frac{1}{i\hbar}$ and  
\begin{equation}
\Sigma^{+}(t=0) = V^{+\dagger}(\fr{\partial S^{+}}{\partial
  \br'},\br') \; .
\end{equation} 

The asymptotic expressions (\ref{eq:GaaAss})-(\ref{eq:GabAss})
are valid as soon as $|\br''-\br'|$ is larger than a few Fermi
wavelengths, which can still correspond to a distance short on the
classical scale, and therefore such that the free Green's function is
a good approximation.  We can therefore use this matching condition to fix the
prefactors $\Sigma^{+}(t=0)$ and $C$ in the generic case, obtaining finally
\begin{equation}
\label{eq:Green}
\begin{split}
& G(\br'',\br';E) = \fr{1}{\sqrt{2i\pi\hbar}} \frac{1}{i\hbar}
\\ 
& \sum_{j} \fr{\exp \left(\fr{i}{\hbar}S_j^{\pm} -
    i\int_{0}^{t_j} M_j^{\pm} dt' -  i\fr{\pi}{2}\mu_j^{\pm}
  \right)}{\sqrt{|J_j^{\pm}|}} 
\, V_j^{\pm}(\br'')\cdot V_j^{\pm\dagger}(\br') \; .  
\end{split}
\end{equation}

\section{Bilayer graphene and Gutzwiller trace formulae}
\label{sec:bilay&GTF}

We turn now to a few extensions of the result derived in
section~\ref{sec:deriv}.  We start with a generalization to the
bilayer graphene case, and then briefly discuss the resulting
Gutzwiller trace formulae for the density of states, valid when
classical periodic orbits are isolated in phase space (i.e.,
generically, for chaotic systems).

\subsection{Semiclassical Green's function for the bilayer case}

The bilayer graphene Hamiltonian can be written at low energy as
\cite{McCann06} 
\begin{equation} \label{eq:Hobi}
{\cal H}^{0}_{bi} = -\fr{1}{2m^{*}}\begin{pmatrix} 0 & (p_x - ip_y)^2
  \\  (p_x + ip_y)^2 & 0 \\ \end{pmatrix} 
\end{equation}
with $m^{*}={\gamma_1}/({2\vf^2})$,  where $\gamma_1$ is the
intra-layer coupling parameter. As before, we would like to include
electric or magnetic fields, as well as a possibly position dependent
mass term.   We therefore  consider the more general Hamiltonian 
\begin{equation}
\label{eq:hambi}
{\cal H}_{bi} = U(\br)\Identity_2 + m(\br)\vf^2\sigma_z + {\cal
  H}^{0}_{bi}(\bp \to {\bf \Pi}) \; . 
\end{equation}

Following the same approach as above, one obtains the semiclassical
Green's function in the form
Eqs.~(\ref{eq:semiGr})-(\ref{eq:Greensemi}) 
except for a different expression of the classical Hamiltonian
eigenenergies 
\begin{equation*}
H^{\pm} = U(\br) \pm \sqrt{m(\br)^2\vf^4 + 
\left( \fr{{\bf \Pi}^2}{2m^{*}} \right)^2}
\end{equation*}
and of the semiclassical (``Berry-like'')  phase term
\begin{equation} \label{eq:BerryBi}
\begin{split} 
M^{\pm} & = \fr{1}{m^{*}}\sqrt{1-\fr{m(\br'')^2\vf^4}{(E-U(\br''))^2}}\\
&\left( \pm e{\bf B} + \fr{1}{2}\fr{{\bf \Pi}\times {\partial
      [m(\br'')\vf^2-U(\br'')]}/{\partial \br''}}{m(\br'')\vf^2+E-U(\br'')}
\right) .{\bf e}_z \; .
\end{split} 
\end{equation}

In the free case ($m(\br) \equiv m_0$, $U(\br) \equiv U_0$), the exact
Green's function can be shown to behave asymptotically as $|\br'' -
\br'| \to +\infty$ as
\begin{eqnarray*}
  G_{AA} & \simeq & 
  \fr{-im^{*}}{4{\hbar}^2} \sqrt{\fr{m_0\vf^2+E-U_0}
    {-m_0\vf^2+E-U_0}} \sqrt{\fr{2}{\pi}}\fr{e^{i(k|\br''- \br'| -
      \fr{\pi}{4})}}{\sqrt{k|\br'' - \br'|}}   \\
G_{\tilde{B}\tilde{B}} & = & G_{AA}(m_0 \to -m_0) \\
G_{A\tilde{B}} & \simeq & \fr{im^{*}}{4{\hbar}^2}e^{-2i\phi}
\sqrt{\fr{2}{\pi}}\fr{e^{i(k|\br'' - \br'| -
    \fr{\pi}{4})}}{\sqrt{k|\br'' - \br'|}} \\
G_{\tilde{B}A} & = & G_{A\tilde{B}}(\phi \to -\phi)
\end{eqnarray*}
with $\phi$ the phase of $p_x + ip_y$.
Matching the exact solution near the source to the semiclassical
expression far from the source eventually gives the semiclassical
Green's function as a sum over all trajectories $j$ joining $\br'$ to
$\br''$ under the classical Hamiltonian $H^+$ or $H^-$ (depending on
the sign of $(E-U(\br'))$)
\begin{equation}
\label{eq:Greenbi}
\begin{split} 
G(\br'',&\br';E) =  \fr{1}{\sqrt{2i\pi\hbar}} \fr{1}{i\hbar} \sum_{j}
V_j^{\pm}(\br'') 
   V_j^{\pm\dagger} (\br') \\
&\frac{\exp\left( {\fr{i}{\hbar}S_j^{\pm} -
    i\int_{0}^{t_j}M_j^{\pm}dt' -
    i\fr{\pi}{2}\mu_j^{\pm}}
  \right)}{\sqrt{|J_j^{\pm}|}} \; ,
\end{split} 
\end{equation}
with $J^\pm$ given by Eq.~(\ref{eq:J}).

\subsection{Trace formulae for isolated orbits}
\label{sec:trace}

One important application of the semiclassical expressions for the
Green's functions is that, by taking their trace, one obtains a
semiclassical approximation for the density of states $\rho(E) =
\sum_i \delta(E-E_i)$.  We have in mind here a quantum dot defined in
a finite region of a graphene sheet (with the confinement imposed for
instance through the mass term), and the $E_i$ are the corresponding
discrete energies of the confined system.  We will furthermore assume
in this subsection the classical motion within the dot fully chaotic,
so that all trajectories are isolated.

Starting from  Eqs.~(\ref{eq:Green}) or (\ref{eq:Greenbi}), the
semiclassical density of states can be obtained as the trace
\begin{equation}
\label{eq:trGr}
\rho(E) \equiv -\frac{1}{\pi} {\rm Im} \int d\br {\rm Tr}[G(\br,\br;E)] \; ,
\end{equation} 
(where $\rm Tr$ is the trace on the internal structure of the Green's
function).
The smooth (Weyl) part of the density of states, which is  associated with
``zero length'' orbits, has the usual expression $\rho_{\rm Weyl}(E) =
\rho^+_{\rm Weyl}(E) + \rho^-_{\rm Weyl}(E) $ with 
\begin{equation*}
\rho^\pm_{\rm Weyl}(E) = \int \fr{d{\bf p}d{\bf r}}{(2\pi\hbar)^2}
\delta(E - H^{\pm}({\bf p},{\bf r})) \; .
\end{equation*} 
When 
potential and mass terms are constant this gives 
\begin{equation} \label{eq:weyl} \rho^\pm_{\rm
    Weyl}(E)=\fr{|E-U_0|\Area}{2\pi(\hbar v_F)^2} \, \Theta \left( \pm(E-U_0) -
  m_{0}v_F^2 \right) \; ,
\end{equation}
 with  $\Area$ the area of the graphene sheet and $\Theta$ the
 Heaviside step function.

The oscillating part $\rho_{\rm osc}(E)$ of the density of states can
then be obtained inserting 
the semiclassical expression for the Green's function in Eq.~(\ref{eq:trGr}).
 Performing the integral on $\br$ in the stationary phase
approximation imposes that, in the semiclassical sums
Eqs.~(\ref{eq:Green}) or (\ref{eq:Greenbi}), only the trajectories
with identical initial and final momentum should be kept. As a
consequence, the sum over the index $j$ becomes a sum over periodic
orbits.  In particular, in Eqs.~(\ref{eq:Green}) or
(\ref{eq:Greenbi}), $V^\pm_j(\br'') = V^\pm_j(\br')$ since
$\br''=\br'=\br$ {\em and} $\bp''_j = \bp'_j$ (remember that
$V^\pm_j(\br) \equiv V^\pm_j(\bp_j(\br),\br)$, so the second condition
is necessary here).  Therefore ${\rm Tr} [V^{\pm }_j(\br'') \cdot
V^{\pm\dagger}_j(\br')]_{|\br''=\br'=\br} = 1$.  Once this point is recognized,
the calculation of $\rho_{\rm osc}$ from the semiclassical Green's functions is,
up to the inclusion of the  semiclassical ``Berry-like''  phase term
$\oint_j M^\pm(t) dt$, essentially the same as in the scalar case
\cite{Gutzwiller71,Gutzwiller90} (see also the particularly clear
discussion in \cite{Creagh90}).  We thus just  quote the final
results:  $\rho(E) = \rho^+(E) + \rho^-(E)$; $\rho^\pm(E) =
\rho^\pm_{\rm Weyl}(E) + \rho^\pm_{\rm osc}(E)$,  with
\begin{equation}
\label{eq:traceGr}
\begin{split}
\rho^\pm_{\rm osc}(E) 
= & \fr{1}{\pi\hbar}\sum_{\rm p.o.}\fr{T_{\rm ppo}}{\sqrt{|{\rm det}
    (\tilde{M_{\rm po}}-1)|}} \\
& \cos{\left(
    \fr{S_{\rm po}^{\pm}}{\hbar}-\fr{\pi}{2}
    \sigma_{\rm po}^{\pm}-\int_{0}^{T_{\rm po}} M^{\pm}dt' \right) }
\; .
\end{split}
\end{equation}
Here $\tilde M = \frac{\partial (p''_\bot,r''_\bot)}{\partial
  (p'_\bot,r'_\bot)} $ is the monodromy matrix, $\sigma^{\pm} =
\mu^{\pm} + \nu^{\pm}$ is the topologically invariant Maslov index
($\nu = 0$ or $1$, depending on the sign of $d^2 S_j / dr_\bot^2$, see
the discussion in \cite{Creagh90}), and $T_{\rm ppo}$ is the period of
the primitive orbit ($T_{\rm po} = n T_{\rm ppo}$ if the orbit consists
of $n$ repetitions of the same path).

\section{Graphene in a constant magnetic field}
\label{sec:landau}

As an illustration of the semiclassical Green's function formalism, we consider
in this section the simple (but useful) case of a graphene sheet
immersed in a constant magnetic field, and show how some standard (and
less standard) expressions can be easily re-obtained in this way.  We start
with the Landau levels in the monolayer and the bilayer, without
potential or mass term ($U(\br)=m(\br)=0$), and assuming the low-energy
approximations  Eqs.~(\ref{eq:Hog})-(\ref{eq:Hobi}) of the Hamiltonian
apply.  We then study the influence of higher order corrections (e.g.
trigonal warping) to this low-energy Hamiltonian.  We finally consider
the case where a finite mass term $m(\br)= m_0 = {\rm const.}$, is
introduced.  This last example  will be used to introduce the
discussion on the distinction between the semiclassical and adiabatic
Berry phases, with which  we shall end this paper in  the next section.

\subsection{Landau levels in monolayer graphene}
\label{sec:LandauMono}

In the absence of confining potential or mass term, and with a
constant magnetic field, the classical equations of motion in graphene
are integrable and lead to cyclotronic motion, i.e. circular periodic
orbits with period $T$ and radius $R$ given in the monolayer case by
\begin{eqnarray} 
T & = & \fr{2 \pi}{\vf^2}\frac{E}{eB} \\
R & = & \fr{\vf}{2\pi} T \; .
\end{eqnarray}
Since the periodic orbits are not isolated, we cannot use the
Gutzwiller trace formula derived in the previous section and we have
to obtain the density of states directly from inserting the
semiclassical expression Eq.~(\ref{eq:Green}) in Eq.~(\ref{eq:trGr}).
Here however the classical dynamics is extremely simple: there is only
one primitive orbit, and the sum over $j$ is actually a sum over the
number of repetitions of this primitive circular orbit.  We therefore
have $S_j^{\pm} = {E}t_j/2$, with $t_j=jT$.  Two caustics are
furthermore traversed for each iteration of the orbit, one midway through the
circle, the other when the orbit comes back to its starting point,
and the Maslov index is thus $\mu_j^{\pm} = 2j$ (note that, as
discussed below, the last caustic should be included).  Finally, the
semiclassical ``Berry-like'' phase term Eq.~(\ref{eq:Mmono}) reduces here
to $M_j^{\pm}(\br(t)) = {{\alpha}\vf^2 e B}/({2E}) = {\rm const.}$ so
that
\begin{equation} \label{eq:BerryMono}
\oint_0^{t_j} M^\pm_j(t) dt = \alpha j \pi \; .
\end{equation}
  
The only technical point in this calculation is therefore that since,
whatever the initial momentum, all trajectories initiated in $\br' =
\br$ eventually return there, the final point $\br'' = \br$ is a
caustic ($\partial r''_\bot / \partial p'_\bot = 0 $) and the prefactor
$1/\sqrt{|J_j|}$ diverges.  As discussed in the appendix~E of
\cite{Richter96}, this divergence can be cured using a mixed
representation of the Green's function, i.e. by expressing the Green's
function $G(\br'',\br')$ in terms of its Fourier transform $\tilde
G(p_x'',y'';x',y')$ as
\begin{equation} \label{eq:FT}
\begin{split} 
G(&x'',y''; x',y') = \\
& \frac{1}{\sqrt{-2i\pi\hbar}} \int dp''_x  \tilde
G(p_x'',y'';x',y') \exp(\frac{i}{\hbar} x'' p''_x) \; .
\end{split} 
\end{equation}
A semiclassical expression for $\tilde G$ can be derived in exactly
the same way as for $G$, and leads to the same expression except for
the transformations $S_j \to \tilde S_j = S_j - p''_x x''$ and $J_j= -
{\dot y}'' {\dot y}' (\frac{\partial^2 S_j}{\partial x'' \partial
  x'})^{-1} \to \tilde J_j = - {\dot y}'' {\dot y}' (\frac{\partial^2
  \tilde S_j}{\partial p''_x \partial x'})^{-1}$.  Thus 
\begin{equation}
\tilde J_j =  {\dot y}'' {\dot y}'
(\frac{\partial  p''_x}{\partial p_x'}) \; ,
\end{equation}
which is not diverging since for the cyclotron motion ${\partial
 p''_x}/{\partial p_x'} = 1$.  The integral over $p''_x$ in
(\ref{eq:FT}) becomes then straightforward (noting that
$dp''_x/{\dot y''} = d\theta$, with $\theta$ the angle made by the
initial velocity with the $x$ axis, this integral basically provides a
factor $\int_0^{2\pi} d \theta = 2 \pi$).  Furthermore the integration over
position in Eq.~(\ref{eq:trGr}) amounts to a multiplication by the
area $\Area$ of the graphene sheet, and as in section~\ref{sec:trace},
${\rm Tr} [V^{\pm}_j(\br'') \cdot
V^{\pm \dagger}_j(\br')]_{|\br''=\br'=\br} = 1$ since the final and initial
momenta are identical.  One therefore obtains 
\begin{equation}
\label{eq:dos}
\rho^{\rm osc}(E) = \fr{|E|\Area
}{\pi({\hbar}\vf)^2}\sum_{j=1}^{+\infty}\cos{2{\pi}j\fr{E^2}{2{\hbar}eB\vf^2}}
\; .
\end{equation}

The total density of states is then $\rho(E) = \rho_{\rm Weyl}(E) +
\rho^{\rm osc}(E)$ with $\rho_{\rm Weyl}(E)$ the smooth density of
states (which is identical to the one without magnetic field) given by
Eq.~(\ref{eq:weyl}).  Using the Poisson formula, we therefore have
\begin{equation}
\label{eq:LL}
\rho(E) = \fr{\Area }{2{\pi}l_B^2}\sum_{n=-\infty}^{+\infty}\delta(E - E_n)
\end{equation}
with $l_B=\sqrt{{\hbar}/({eB})}$ and 
\begin{equation} \label{eq:EnMono}
E_n={\rm sign}\,(n) \vf\sqrt{|2n{\hbar}eB|} \; .
\end{equation}
We recover in this way the expression of the Landau levels as obtained
in a fully quantal derivation \cite{McClure56}.  This approach
furthermore provides a direct link between the phase $\oint_0^{t_j}
M_j(t) dt = \alpha j \pi$ and the existence of a zero energy level, as
it cancels out the phase associated with the Maslov indices (another
example of such a cancellation can be found in \cite{Keppeler02}).  An
alternative semiclassical derivation of the graphene Landau levels can
be obtained starting from the Dirac oscillator \cite{Keppeler03}, in
the limit of massless carriers, provided the frequency of the
oscillator is taken to be the cyclotronic one.

\subsection{Landau levels in bilayer graphene}

Considering now the bilayer case, we can proceed in exactly the same
way as above except for two differences.  First, the period $T$ and
radius $R$ are now given by 
\begin{eqnarray} 
T & = & \frac{2\pi}{\omega} = {2 \pi} \fr{m^{*}}{eB}\\
R & = &  \sqrt{\fr{|E|}{2\pi^2 m^{*}}} T \; .
\end{eqnarray}
Second, the semiclassical ``Berry-like'' phase term Eq.~(\ref{eq:BerryBi})
now reduces to $M_j^{\pm}(\br(t)) = \pm{eB}/{m^{*}} = {\rm const.}$,
so that
\begin{equation}
\oint_0^{t_j}  M_j(t) \,dt = 2 j \pi \; .
\end{equation}
The Berry-like phase does not in this case compensate the phase
associated with the Maslov index.  Noting furthermore that, for the
bilayer graphene, $\rho_{\rm Weyl}(E) = {m^{*}\Area }/({2\pi{\hbar}^2})$, we obtain
\begin{equation}
\label{eq:LLbi}
\rho(E) = \fr{\Area }{2{\pi}l_B^2}\sum_{n=-\infty}^{+\infty}\delta(E -
E_n^{\rm sc}) \, 
\end{equation}
where
\begin{equation}
E_n^{\rm sc} = \hbar \omega (n-\fr{1}{2}) 
\end{equation}
is the semiclassical approximation to the exact quantum values of the
Landau levels, ${E_n}^{\rm quant} = \hbar \omega \sqrt{n(n-1)} = \hbar
\omega (n-\fr{1}{2})+O(\fr{1}{n})$.  The semiclassical calculation
fails here to account for the $O(\fr{1}{n})$ term.  The $n=0$ and
$n=1$ Landau levels, which both have zero energy, are therefore not
correctly described within this semiclassical approach. However, for
$n \ge 2$, the agreement between the semiclassical approximation and the
exact result is quantitatively very good. 


\subsection{Influence of higher order corrections (in the parameter
  $(a |\bp| /\hbar)$)}
\label{sec:trigonal}

The next example to which we shall apply our semiclassical formalism
is the shift of the Landau levels associated with deviations, for
large momenta, to the linear approximation of the graphene dispersion
relation Eq.~(\ref{eq:Hog}) \cite{Plochocka07}.

Starting from a tight-binding description of the graphene monolayer in
which the effect of the next-to-nearest neighbor hopping is taken into
account via the parameter $t' \ll t$, and expanding the resulting
dispersion relation near the $\bK$ and $\bK'$ points up to third order
in $(a |\bp| /\hbar)$ (the reason for expanding up to third order will
become clear below), the resulting Hamiltonian reads (in the absence
of electric or magnetic fields) \cite{Plochocka07}
\begin{equation}
\label{eq:ham2}
{\cal H}'_g = {\cal H}_g^0 + \begin{pmatrix} h'(\bp) & h(\bp)^{*}
  \vspace{0.2cm} \\ h(\bp) & h'(\bp) \\ \end{pmatrix} 
\end{equation}
with ${\cal H}_g^0$ given by Eq.~(\ref{eq:Hog}) and 
\begin{eqnarray*}
  h'(\bp) & = & -3t' + 6\fr{t'}{t}\vf|\bp| \left( \fr{\vf}{6t}|\bp| -
   2\alpha(\fr{\vf}{6t})^2 {\bp}^2 \cos{3\phi_{\bp}} \right) \\
h(\bp) & = & -\vf \left( \fr{\vf}{6t}({\alpha}p_x - ip_y)^2 +
  2(\fr{\vf}{6t})^2{\bp}^2({\alpha}p_x+ip_y) \right) \; .
\end{eqnarray*}
Keeping only terms no greater than third order in momentum, the
eigenvalues of the associated classical Hamiltonian can be
expressed as
\begin{equation}
\begin{split}
H^{\pm}= h'(\bp) \; {\pm} \; \vf|\bp|& \left(1-\alpha\fr{\vf}{6t}|\bp|
\cos{3\phi_{\bp}}  \right.\\
- & \left. \fr{1}{2} (\fr{\vf}{6t})^2
{\bp}^2(3+\cos^2{3\phi_{\bp}}) \right) \; ,
\end{split}
\end{equation}
with $\phi_{\bp}=\arctan({p_y}/{p_x})$. The anisotropic terms,
proportional to $\cos{3\phi_{\bp}}$, are often referred to as trigonal
warping. Recall now this expansion is valid if the condition $|\bp|
\ll {\hbar}/{a}$ is fulfilled. Rewriting the expression
$({\vf}/{6t})|\bp| = {|\bp|a}/({4\hbar})$, higher order terms in
$H^{\pm}$ can thus be viewed as a perturbation of the original
eigenvalue $H^{\pm}={\pm}\vf|\bp|$ in the small parameter $(\lambda
\vf |\bp|)$, where $\lambda \equiv {\alpha}/{6t}$ will be used below
to identify the order in the perturbation.  In the semiclassical limit
($\hbar \to 0$), only the modification of the action needs to be taken
into account since this latter is multiplied by the large parameter
$1/\hbar$.  Our aim is therefore to compute the (first and second order here)
corrections to the action in an expansion in $\lambda$
\begin{equation}
\label{eq:action}
S = S_0 + \lambda\delta^{(1)}S + {\lambda}^2\delta^{(2)}S \; .
\end{equation}

In the presence of a constant magnetic field, the classical equations
of motion derived from the first order approximation $H^{\pm}=\pm \vf|{\bf
  \Pi}|$ are integrable, and this property is not modified by the
addition of terms in $H^\pm$ depending only on $|{\bf \Pi}|$. This can
be easily shown by performing a canonical transformation to the guiding
center coordinates.  For sake of completeness, this canonical change of
variables is detailed in appendix~\ref{sec:appB}. The new coordinates
read
\begin{eqnarray*}
{\bf R} & = & (\fr{1}{eB}\Pi_y,x_0) \\
{\bf P} & = & (\Pi_x,eBy_0)
\end{eqnarray*}
with $\br_0$ the center of the cyclotron orbit, so that $|{\bf
\Pi}|=\sqrt{P_x^2+(eBX)^2}$ and $\tan{\phi_{\bf
  \Pi}}={eBX}/{P_x}$. We thus have 
\begin{equation*}
H^{+} = -3t' + \rho -\mu_2\lambda\rho^2 -\mu_1{\lambda}^2\rho^3 
\end{equation*}
with $\vf(P_x+ieBX)=\rho e^{i\phi}$,
$\mu_2=(\cos{3\phi}-6\alpha\fr{t'}{t})$, and
$\mu_1=\fr{1}{2}(3+\cos^2{3\phi}+6\alpha\fr{t'}{t}\cos{3\phi})$. In
this new system of coordinates, the 
action is easily calculated as 
\begin{equation*}
S = \int {\bf P}d{\bf R} = \int P_xdX =
\fr{1}{2\vf^2eB}\int_{0}^{2\pi}{\rho^2}(\phi)d\phi 
\end{equation*}
with the constraint $E=H^{+}$.  Therefore, to order ${\lambda}^2$, and
with $E'=E+3t'$
\begin{equation*}
\rho^2 = E'^2 + 2\mu_1{\lambda}E'^3 +
(5{\mu_1}^2+2\mu_2){\lambda}^2E'^4 
\end{equation*} 
which gives for the action
\begin{equation*}
\begin{split}
S = \fr{1}{2\vf^2eB} & \left( 2\pi E'^2 -
  24\pi\alpha\fr{t'}{t}{\lambda}E'^3 \right. \\
+ & \left.  12\pi(1+30(\fr{t'}{t})^2){\lambda}^2E'^4 \right) \; .
\end{split}
\end{equation*}
The third order terms had to be taken into account in the low-energy
expansion, since their contribution in the second order correction of
the action is of the same magnitude as that of second order terms. The
third order term in the next-to-nearest neighbor contribution however
cancels out in the calculation of $S$ and thus a second order
expansion in $h'(\bp)$ would have been sufficient. Introducing this
shift in the action in the Landau-levels  calculation of 
section~\ref{sec:LandauMono} finally gives
\begin{equation}
\label{eq:shift}
\begin{split}
E'_n = E_n \left( 1 \pm 6\alpha\fr{t'}{t}{\lambda}E_n - 
3{\lambda}^2E_n^2 \right)
\\
=E_n \left( 1 \pm \fr{3t'}{\sqrt{2}t}\fr{a}{l_B}\sqrt{n} - 
\fr{3}{8}(\fr{a}{l_B})^2n \right)
\end{split} 
\end{equation}
($l_B=\sqrt{{\hbar}/({eB})}$ is the magnetic length).  This result
is identical to the one obtained purely quantum mechanically in
[\onlinecite{Plochocka07}].  As discussed in this paper, the resulting effect
is however too small to interpret shift in Landau levels observed
experimentally by Plochocka et al. \cite{Plochocka07} .


\subsection{Effect of a mass term}

To end this section, let us consider the effect of a constant  mass
term $m_0 \vf^2 \sigma_z$  in the graphene Hamiltonian, so that 
\begin{equation}
H^\pm = \pm \sqrt{m_0^2 \vf^4 + \vf^2 {\bf \Pi}^2} \; .
\end{equation}
Interestingly, a constant mass term does not modify the time
derivative $M(t)$ of the semiclassical Berry-like phase since (see
Eq.~(\ref{eq:Mmono})) it depends only on the gradient of
$m(r)$. Furthermore,  as shown by a direct calculation,
the energy dependence of the Landau frequency is not affected either
by the mass term. Therefore
\begin{eqnarray}
T & = & \frac{2\pi}{\omega} = \frac{2 \pi}{\vf^2} \fr{E}{eB}\\
M(\br(t)) & = & \alpha \vf^2 \fr{eB}{2E} = {\rm const.} \; ,
\end{eqnarray}
and the semiclassical phase 
\begin{equation} \label{eq:BerryGap}
\oint_0^{t_j} M_j(t) dt = j \alpha \pi
\end{equation}
is the same as without the mass term.

The $m_0$ dependence of the Landau level position is therefore
entirely due to the $m_0$ variation of the action
\begin{equation}
  S_j = j\pi \frac{E^2}{eB \vf^2} \left[1 - \left(\frac{m_0
        \vf^2}{E} \right)^2 \right] \; ,
\end{equation}
which, following the same steps as in section~\ref{sec:LandauMono},
gives $ \rho(E) = ({\Area}/({2\pi l_B^2})) \sum_{n=0}^{+\infty}
 \delta(E \pm E_n)$, with 
\begin{eqnarray}
E_n & = &  \sqrt{E^2_n(0)  + m_0^2 \vf^4} \\
& \simeq & E_n(0) \left( 1 + \frac{1}{4n}
  \frac{(m_0 \vf)^2}{e\hbar B} \right) 
\end{eqnarray}
($E_n(0)$ is the value of $E_n$ at $m_0=0$ given by
Eq.~(\ref{eq:EnMono})).  One recovers semiclassically in this way the
result originally derived by Haldane \cite{Haldane88}.

\section{Semiclassical versus adiabatic Berry phase}
\label{sec:1/2classvsquant}

We would like to finish this paper with some general discussion
concerning the semiclassical phase 
\begin{eqnarray} 
\xi_{\rm sc} & \equiv & - \oint_0^T M_{\rm sc}(\bp(t),\br(t)) dt \\
M_{\rm sc}(\bp(t),\br(t)) & = & {\rm Im} \left[
  V^{\pm\dagger}\fr{\partial H}{\partial 
  \bp}. \left( \fr{\partial V^{\pm}}{\partial \br} \right) \right] 
\label{eq:Msc}
\end{eqnarray}
(see Eq.~(\ref{eq:Box2})) computed on a periodic orbit $(\bp(t),\br(t))$
(of period $T$).

That, for a clean graphene monolayer without a mass term, $\xi_{\rm
  sc} = \mp \pi$ (as expressed by Eq.~(\ref{eq:BerryMono}), with
$j=1$) is usually said to be expected since the corresponding
configuration is exactly the one discussed in detail by Berry in his
1984 paper \cite{Berry84}: the path of integration corresponds to
encircling once the Dirac point, where the $H^+$ and $H^-$ manifolds
intersect.  This argument however relies on an exact intersection
between the two manifolds, and should a priori not apply when a mass
term $m_0$ introduces a gap.  From this perspective, one does not
expect the Berry phase to be equal to $\pm\pi$ when $m_0 \neq 0$, and
Eq.~(\ref{eq:BerryGap}) may come as a surprise. (Note though this was
already observed in \cite{Gusynin06}.)

The resolution of this apparent paradox is that, as discussed in
[\onlinecite{Littlejohn91prl,Littlejohn91pra}], the semiclassical
phase $\xi_{\rm sc}$ defined by Eq.~(\ref{eq:gammasc}) and the
adiabatic phase introduced by Berry are closely related, but
eventually different, quantities.  Both of them are induced by the
adiabatic variation of the eigenstates $V^+$ and $V^-$ along the
trajectory.  However, the point of view taken in the semiclassical
approach is that both the internal space (associated here with the
sub-lattices $(A,B)$) and the external space (position $\br$) are {\em
  coupled} dynamical variables.  Treating the coupling between these
variables in the semiclassical approximation (which indeed implicitly
assumes that the ``external'' variable is slow and the internal
variable fast) leads to the semiclassical expression (\ref{eq:Msc}).

The problem Berry was considering in his seminal article
\cite{Berry84} is however different: in that case, only the internal
degree of freedom is considered a dynamical variable, and the
external degrees of freedom are actually a space of parameters assumed
to be entirely controlled by the experimentalist.  One may in that case
of course choose this path as the classical trajectory $(\br(t))$ (with
$H(\br) \equiv H(\bp(\br),\br)$) determined by the dynamics in the
semiclassical approach.  In that case however the corresponding phase
is given by \cite{Berry84}
\begin{eqnarray} 
\xi_{\rm ad} & \equiv & \oint_0^T M_{\rm ad}(\br(t)) dt \\
M_{\rm ad}(\br(t)) & = & i   V^{\pm\dagger} \frac{\partial
  V^{\pm}}{\partial \br} \cdot \dot\br
\label{eq:Mad}
\end{eqnarray}
(the normalization of $V^\pm$ ensures that $M_{\rm ad}$ is real).

Let us assume, for this discussion, that we are interested in the
evolution of the eigenstate $V^+$ associated with the positive
eigenvalue $H^+$.  Furthermore, let us switch to the bra/ket notation
for the eigenvector and write $V^\pm \equiv | \pm \rangle$, $V^{\pm
  \dagger} \equiv \langle \pm |$.  First order perturbation theory
implies $\dot \br =
\partial H^+ / \partial \bp = \langle \smpl |(\partial H / \partial \bp)
| \smpl \rangle$, and therefore Eq.~(\ref{eq:Mad}) can be rewritten as 
\begin{equation} 
M_{\rm ad}(\br(t)) =  i  \langle \smpl | \frac{\partial H}{\partial
  \bp}| \smpl \rangle \cdot  \langle \smpl | \frac{\partial }{\partial \br}\smpl
\rangle  \; .
\end{equation}
On the other hand, inserting the identity $ \Identityy_2 = | \smpl \rangle
\langle \smpl | + | \smmi \rangle \langle \smmi |$ in Eq.~(\ref{eq:Msc}),
\begin{equation} 
M_{\rm sc}(\br(t)) =  {\rm Im} \left[ \langle \smpl | \frac{\partial H}{\partial
  \bp}| \smpl \rangle \cdot  \langle \smpl | \frac{\partial }{\partial \br}\smpl
\rangle  
+
 \langle \smpl | \frac{\partial H}{\partial
  \bp}| \smmi \rangle \cdot  \langle \smmi | 
\frac{\partial }{\partial \br}\smpl
\rangle  \right] \; .
\end{equation}
Thus, the adiabatic and semiclassical phases actually differ
from the quantity
\begin{equation}
\xi_{\rm ad} - \xi_{\rm sc} = \oint_0^T  {\rm Im} \left[ \langle \smpl
  | \frac{\partial H}{\partial 
  \bp}| \smmi \rangle \cdot  \langle \smmi | 
\frac{\partial }{\partial \br}\smpl
\rangle  \right]   dt \; .
\end{equation}

From this expression, we can see that {\em in the absence of a mass
  term}, but for an arbitrary electrostatic potential $U(\br)$, the
semiclassical and Berry phases are identical.  Indeed, for $m(\br)
\equiv 0$, the expressions Eqs.~(\ref{eq:V+})-(\ref{eq:V-}) for  the
eigenvectors of $H(\bp,\br)$ take the simple form
\begin{eqnarray}
| \smpl\rangle & = & \fr{1}{\sqrt{2}} 
\begin{pmatrix} 1 \\ \alpha e^{i\alpha\phi} \end{pmatrix}
\label{eq:V0+}\\
| \smmi \rangle & = & \fr{1}{\sqrt{2}} 
\begin{pmatrix} \alpha e^{-i\alpha\phi} \\ -1 \end{pmatrix} \; ,
\label{eq:V0-}
\end{eqnarray}
with $\phi$ the phase of $\Pi_x + i \Pi_y$.  As a consequence
\begin{eqnarray} 
\langle \smpl   | \frac{\partial H}{\partial  \bp}| \smmi \rangle
\cdot  \langle \smmi | \frac{\partial }{\partial \br}\smpl \rangle  
& = & \frac{\vf}{2} (- \sin \phi  \partial_x \phi + \cos \phi
\partial_y \phi)  \label{eq:pm}\\
\langle \smpl   | \frac{\partial H}{\partial  \bp}| \smpl \rangle
\cdot  \langle \smpl | \frac{\partial }{\partial \br}\smpl \rangle  
& = & i \frac{\alpha\vf}{2} (  \cos \phi  \partial_x \phi + \sin \phi
\partial_y \phi) \nonumber \\
& = & \frac{i\alpha}{2} \frac{d \phi}{dt} \; .\label{eq:pp}
\end{eqnarray}
The right hand side of Eq.~(\ref{eq:pm}) is purely real, implying
that, in the simple case $m=0$ considered here, $\xi_{\rm ad} -
\xi_{\rm sc} = 0$.  Eq.~(\ref{eq:pp}) then expresses that,
independently of the nature of the electrostatic potential $U(\br)$,
the -- here identical -- Berry phase and semiclassical phase are just
given by plus or minus (depending on $\alpha$) half the angle of
rotation of the velocity vector.  In particular, as demonstrated by
Berry from geometric arguments \cite{Berry84}, we see here from a
direct calculation that for a periodic orbit, $\xi_{\rm ad} = \xi_{\rm
  sc} = -\alpha j \pi$, with $j$ the number of windings of the trajectory.
This makes particularly simple  the inclusion of the semiclassical
phase in the Gutzwiller trace formula Eq.~(\ref{eq:traceGr}) when
$m=0$.

Similarly, for the bilayer Hamiltonian Eq.~(\ref{eq:hambi}) with
$m(\br) \equiv 0$, we have 
\begin{eqnarray}
| \smpl\rangle & = & \fr{1}{\sqrt{2}} 
\begin{pmatrix} 1 \\- e^{i2\phi} \end{pmatrix}
\label{eq:V0+bi}\\
| \smmi \rangle & = & \fr{1}{\sqrt{2}} 
\begin{pmatrix} e^{-i2\phi} \\ 1 \end{pmatrix} \; ,
\label{eq:V0-bi}
\end{eqnarray}
with $\phi$ the phase of $\Pi_x + i \Pi_y$, and
\begin{eqnarray} 
\langle \smpl   | \frac{\partial H}{\partial  \bp}| \smmi \rangle
\cdot  \langle \smmi | \frac{\partial }{\partial \br}\smpl \rangle  
& = & \frac{|\Pi|}{m^*} (- \sin \phi  \partial_x \phi + \cos \phi
\partial_y \phi)  \label{eq:pmbi}\\
\langle \smpl   | \frac{\partial H}{\partial  \bp}| \smpl \rangle
\cdot  \langle \smpl | \frac{\partial }{\partial \br}\smpl \rangle  
& = & i \frac{|\Pi|}{m^*} (  \cos \phi  \partial_x \phi + \sin \phi
\partial_y \phi) \nonumber \\
& = & i \frac{d \phi}{dt} \; .\label{eq:ppbi}
\end{eqnarray}
Again, the Berry phase and semiclassical phase are identical if
$m(\br) \equiv 0$ (as Eq.~(\ref{eq:pmbi}) is purely real), and both
phases are given by the angle of rotation of the velocity vector.

For both bilayer and monolayer graphene, it has to be born in mind
however that in the generic case $m(\br) \neq 0$, the semiclassical
phase $\xi_{\rm sc}$ should in general differ from the Berry phase
$\xi_{\rm ad}$.  Furthermore, we do not have a general argument
constraining any of the two phases to be directly related to the
winding of the velocity vector (beyond the case where both the mass
and the electrostatic potential are constant).

\section{Conclusion}

To conclude, we have derived an expression for the semiclassical
Green's function in graphene and discussed in particular the
semiclassical phase associated with the internal pseudo-spin
structure.  If no mass term is included in the graphene Hamiltonian,
this semiclassical phase is identical to the corresponding (adiabatic)
Berry phase.  In that case both phases are, up to a sign, given by
half the angle of rotation of the velocity vector.  For a bilayer of
graphene, the same result holds but with a phase which is twice as
large.

When a mass term is introduced however, the semiclassical and Berry
phases in general differ. In particular, for a clean graphene sheet in
a constant magnetic field, we have shown that the semiclassical phase
remains unmodified upon the inclusion of a constant mass term $m (\br)
= m_0$, while the corresponding Berry phase $\xi_{\rm ad} =
[m_0/(E-U_0)-1] \alpha j \pi$ shows some dependence on
$m_0$.  We have shown furthermore that in this case, what is relevant
to the calculation of the Landau levels is the semiclassical, rather
than the Berry, phase.
Other applications of our semiclassical formalism were also discussed,
including the effect of higher order terms of the graphene Hamiltonian
-- e.g.\ trigonal warping -- on the position of the Landau levels.

The semiclassical approximation to the graphene Green's function
should prove a useful tool when considering confined electron
systems in graphene, such as graphene nanoribbons, or more complicated
geometries.

We have benefited from helpful discussions with E. Bogomolny, J.-N.
Fuchs, M.-O. Goerbig, G. Montambaux and F. Pi\'echon, and thank as
well all active participants of the weekly graphene ``journal-club''
held in LPS, Orsay.

\appendix

\section{Imaginary part of the operator ($\Box_2$)}
\label{sec:appA}

In this appendix, we give the details of the computation of the
imaginary part ${\rm Im}(\Box_{(2)}) = M^{\pm}$ (see
eq.~(\ref{eq:BK})) of the operator
\begin{equation}
\Box_{(2)} = \vf V^{\pm\dagger} \cdot
\left( \alpha\sigma_x\fr{\partial V^{\pm}}{\partial x} +
  \sigma_y\fr{\partial V^{\pm}}{\partial y} \right) \; .
\end{equation}
Here
\begin{eqnarray}
V^{+}(\bp,\br) & = & 
\fr{\begin{pmatrix} 
m(\br) \vf^2 + \epsilon(\bp,\br) \\ 
\vf (\alpha \Pi_x + i \Pi_y)  
\end{pmatrix}}
{\sqrt{2(\epsilon(\bp,\br))(m(\br)\vf^2+\epsilon(\bp,\br))}} 
\label{eq:V+}\\
V^{-}(\bp,\br) &=&  
\fr{\begin{pmatrix} 
   \vf(\alpha\Pi_x - i\Pi_y) \\
    -(m(\br)\vf^2+\epsilon(\bp,\br)) \\
\end{pmatrix}}{\sqrt{2(\epsilon(\bp,\br))(m(\br)\vf^2+\epsilon(\bp,\br))}} 
\label{eq:V-}
\end{eqnarray}
are the normalized eigenvectors of the classical Hamiltonian $H^\pm$ (see
Eq.~(\ref{eq:Hclass})),  
\begin{equation*}
\epsilon(\bp,\br) = H^{+}(\bp,\br)-U(\br) = \sqrt{m(\br)^2\vf^4 + \vf^2{\bf \Pi}^2}
\end{equation*}
and, with respect to spatial derivation, it is understood that
$V^\pm \equiv V^\pm[({\partial S^{\pm}}/{\partial \br}),\br]$.  

We perform here the calculation for $M^{+}$, the one for $M^{-}$ being
essentially identical.  We have
\begin{widetext}
\begin{eqnarray}
\alpha  {\rm Im}  \left[ V^{+\dagger} \sigma_x (\partial_x V^{+}) \right] 
&  = &  \fr{\alpha  }{{2\epsilon(m\vf^2+\epsilon)}}  
{\rm Im} \left[ 
\left( m \vf^2 + \epsilon , \vf (\alpha \Pi_x - i \Pi_y) \right)
\cdot \sigma_x \cdot
\begin{pmatrix} 
  \partial_x (m \vf^2 + \epsilon) \\
  \vf \partial_x (\alpha \Pi_x + i \Pi_y) 
\end{pmatrix}
\right]  
\nonumber \\
& = & \fr{\alpha \vf }{2 \epsilon (m \vf^2 + \epsilon)}
\left[ (m \vf^2 + \epsilon) \partial_x \Pi_y
- \Pi_y \partial_x (m\vf^2+\epsilon) \right] \; ,
\end{eqnarray}
and in the same way
\begin{eqnarray}
 {\rm Im}  \left[ V^{+\dagger} \sigma_y (\partial_y V^{+}) \right] 
&  = &  \fr{1 }{2\epsilon (m \vf^2 + \epsilon)}  
{\rm Im} \left[ 
\left( m \vf^2 + \epsilon  ,   \vf (\alpha \Pi_x - i \Pi_y) \right)  
\cdot \sigma_y  \cdot
\begin{pmatrix} 
{\partial_y (m\vf^2+\epsilon)} \\
  \vf \partial_y (\alpha \Pi_x + i \Pi_y) 
\end{pmatrix}
\right] \nonumber \\
& = & \fr{\alpha \vf}{{2\epsilon(m\vf^2+\epsilon)}}
\left[ (m \vf^2 + \epsilon) (-\partial_y \Pi_x)
+ \Pi_x \partial_y (m\vf^2+\epsilon) \right] \; ,
\end{eqnarray}
\end{widetext}
so that 
\begin{equation*}
{\rm Im}(\Box_{(2)}) = \fr{\alpha \vf^2}{2\epsilon} \left(
  \fr{\partial }{\partial \br} \times {\bf \Pi} + \fr{{\bf \Pi}\times
    {\partial_\br (m\vf^2+\epsilon)}}{m\vf^2+\epsilon}
\right).{\bf e}_z \; ,
\end{equation*}
with ${\bf e}_z $ the unit vector in the direction perpendicular to
the graphene plane.

Using finally that 
\begin{eqnarray*}
\left( \fr{\partial }{\partial \br} \times {\bf A} \right)_z & =  & B
\\
\left( {\partial_\br} \times {\partial_\br S } \right)_z & =  & 
\partial_x\partial_y S - \partial_y\partial_x S = 0
\end{eqnarray*}
and that the Hamilton-Jacobi equation $E-H^{+}(\fr{\partial
  S^{+}}{\partial \br},\br)=0  $ implies 
\begin{equation*}
\fr{\partial }{\partial \br} \left( \epsilon(\fr{\partial
    S^{+}}{\partial \br},\br) \right) = -\fr{\partial U}{\partial \br} 
\end{equation*}
then gives Eq.~(\ref{eq:Mmono}).

\section{Guiding center coordinates}
\label{sec:appB}

We sketch here the construction of the new canonical variables $({\bf
  R}=(X,Y),{\bf P}=(P_x,P_y))$ introduced in section~\ref{sec:trigonal}.  We start
first by performing the simple canonical transformation $\br \to \br'
= (x,p_y), \bp \to \bp' = (p_x,-y)$.  Then, introducing the guiding
center $\br_0 = (x_0,y_0)$ coordinates
\begin{eqnarray*}
x_0 & = &  x - \fr{1}{eB}\Pi_y = \fr{x}{2} - \fr{1}{eB}p_y \\
y_0 & = & y + \fr{1}{eB}\Pi_x = \fr{y}{2} + \fr{1}{eB}p_x
\end{eqnarray*}
we define the point transformation ${\bf R}(\br')$ as
\begin{equation*}
{\bf R} = (\fr{x}{2} + \fr{1}{eB}p_y, \fr{x}{2} - \fr{1}{eB}p_y) =
(\fr{1}{eB}\Pi_y, x_0)  \; .
\end{equation*}
This transformation is obtained from the generating function
$F(\br',{\bf P}) = {\bf P}.{\bf R}(\br')$, and therefore the new
momentum is given by  
\begin{equation*}
\bp' = \fr{\partial F}{\partial \br'} = \begin{pmatrix} \fr{1}{2}(P_x
  + P_y) \vspace{0.2cm} \\ \fr{1}{eB}(P_x - P_y) \\ \end{pmatrix} 
\end{equation*}
which is easily inverted into 
\begin{equation*}
 {\bf P}(\bp') = (p_x - \fr{eB}{2}y, p_x + \fr{eB}{2}y) = (\Pi_x,
 eBy_0)  \; .
\end{equation*}
The unperturbed Hamiltonian is then given as $H^{+}(\bp,\br) =
\vf|{\bf \Pi}| = \vf\sqrt{P_x^2 + (eBX)^2}$.

\bibliography{graphene1}

\end{document}